Individual Performance and Leader's Laterality in Interactive Contests

Satyam Mukherjee[1, 2,*]

[1]Northwestern Institute on Complex Systems (NICO), Northwestern University Evanston

[2]Indian Institute of Management Udaipur, Udaipur City

* Corresponding author (satyam.mukherjee@gmail.com)

Abstract

Left-handedness is known to provide an intrinsic and tactical advantage at top level in many sports involving interactive contests. Again, most of the renowned leaders of the world are known to have been left-handed. Leadership plays an important role in politics, sports and mentorship. In this paper we show that Cricket captains who bat left-handed have a strategic advantage over the right-handed captains in One Day International (ODI) and Test matches. The present study involving *46* left-handed captains and *148* right-handed captains in ODI matches, reveal a strong relation between leader's laterality and team-member performance, demonstrating the critical importance of left-handedness and successful leadership. The odds for superior batting performance in an ODI match under left-handed captains are *89%* higher than the odds under right-handed captains. Our study shows that left-handed captains are more successful in extracting superior performance from the batsmen and bowlers in ODI and Test matches; perhaps indicating left-handed leaders are better motivators as leaders when compared to right-handed captains.

## 1. INTRODUCTION

It has been widely accepted that 10% of humans are left-handed (Raymond et al 1996). Polymorphism in human handedness existed at least since the Upper Paleolithic and all human populations have displayed significant bias towards right-handedness (Faurie & Raymond 2004). It has been established that left-handers enjoy an advantage in sports, particularly those involving team activity like baseball (Goldstein & Young 1996) and cricket (Wood & Aggleton 1989). Left-handed advantage is also seen in sports with one-to-one interaction like tennis (Holtzen 2000), fencing (Harris 2010), boxing (Gursoy 2008), wrestling (Ziyagil, Gursoy, Dane & Yuksel 2010), and mixed martial arts (Baker & Schorer 2013). Earlier studies have shown that the low percentage of left-handers is a result of the balance between cooperation and competition in human evolution (Abrams & Panaggio 2012; Ghirlanda, Frasnelli & Vallortigara 2009) as well as other vertebrates (Ghirlanda & Vallortigara 2004; Vallortigara 2006). Previous studies have illustrated a positive association between left-handedness and cognitive ability of mathematically talented individuals (Coren 1995). An earlier study conducted on university students revealed a positive association between left-handedness and divergent thinking in male students considered in the study. More recent research indicates that sinistrality is linked with better spatial skills and more working memory (Beratis et. al 2013).

Even though left-handers comprise of 10% of the general population, *6 out of the last 15* U.S Presidents have been left-handed. Presidents Gerald R. Ford, George Bush, Bill Clinton and Barack Obama are all left-handed. The imbalance between general population and left-handers gives rise to the popular belief whether left-handers are efficient leaders (Holder 2001). Leadership is one of the most important parameters to judge one's ability to lead or mentor in any team activities. However, till date there has been no study or evidence to validate the effect of laterality on successful leadership. Even though some left-handers may be better equipped to accept the challenges of leadership in politics, there is no evidence or study to prove whether the left-handed leaders are successful during their tenure. In this work, the potential effects of handedness on leadership are investigated within the premises of sports, a prolific testing ground for laterality research. Left-handed athletes like Babe Ruth (in baseball), Larry Bird (in basketball) and Clive Lloyd (in cricket) are celebrated as great motivators in the world of sports. In this paper we attempt to address the link between left-handedness and leadership taking help of the wealth of data available in sports (Baker & Schorer 2013; Radicchi 2005; Duch, Waitzman, & Amaral 2010; Mukherjee 2012; Saavedra, Powers, McCotter, Porter, & Mucha 2009; Skinner 2010; Saavedra, Mukherjee, & Bagrow 2012).

Cricket is a sport in which left-handers have always enjoyed an advantage (Raymond et al 1996). They are represented more often than expected in the general population. Almost a century ago only three of the top 30 batsmen were left-handers. Only Clem Hill of Australia held the top position. The succeeding decades did not see any major change. However, by 1934 the only significant left-handers were Maurice Leyland and Eddie Paynter (See Table 1). With the advent of Arthur Morris, Neil Harvey, Bert Sutcliffe and, above all, Sir Garry Sobers, left-handers started occupying elite (among top 100) positions during the 1950s. Contrary to the batsmen there is no corresponding increase in left-arm bowlers. In 1904 there were six left-handed players in the top 20, with Wilfred Rhodes occupying second position in International Cricket Council (ICC) rankings. A glossary of Cricket terms used in the paper is included in the Appendix.

Cricket is also a game in which an outcome depends a lot on the leadership. Contrary to the 'off-field' role of coach in Soccer or manager in Baseball, a Cricket captain is directly involved in the proceedings of a game. This can be viewed as team-leadership in the corporate world and leadership in politics. The captain chooses the batting order, sets up fielding positions and shoulders the responsibility of on-field decision-making and is also responsible at all times for ensuring that play is conducted within the Spirit of the Game as well as within the Laws. Compared to the role of coach in Soccer or manager in Baseball, the role of a Cricket captain is not restricted to off-the-field decisions but also to deliver winning performance for the team while playing in the field. In a team game like Cricket a lot of media scrutiny and criticism is showered on the captains. Lots of TV commentators and Cricket experts believe that pressure of leading the teams affect the performance of the players in the team.

In this paper we address the question - Do left-handers make better leaders. Earlier studies on handedness distribution research focused predominantly on data accumulated from several years (Aggleton, Bland, & Kentridge 1994). To understand the impact of handedness on performance it is essential to track the temporal distribution of handedness. For example, in Major League Baseball, frequencies of left-handed pitchers and batters increased logarithmically with time (between 1876 and 1985) and attained a stable overrepresentation at 30% (Wood & Aggleton 1989). Again, it was observed that in 2007 ICC Cricket World Cup left-handed batsmen were more successful than the right-handed batsmen, and that most successful teams had approximately *50.5*% left-handed batsmen (Brooks et al 2004). Again it was found that a higher proportion (*20*%) of professional cricketers bowled with their left hand, leading to the theory that the left-handed advantage is tactical (Wood & Aggleton 1989).

In this paper, first we perform a temporal study of the representation of left-handed batsmen and bowlers in elite category of top 100 performers in history of Cricket and subsequently analyzed whether handedness of the captain play any significant role in the individual performance of players. We specify that leadership is a potential aspect of left-handedness, which is not included in current theoretical models.

## 2. METHODS
*(a) Data*

We extract information of all *3420* One Day International (ODI) matches played between 1971 and 2013 and all *1979* Test matches played between 1877 and 2013 recorded in Cricinfo (Cricinfo 2014). For every match we record the personal performance of batsmen and bowlers under different captains. Data for captaincy records is collected from the Cricinfo website (Cricinfo 2014). In Test cricket out of *304* captains, *54* of them bat left-handed. Again in ODI cricket we collect information of *194* captains, of which *46* are left-handed. Data for international Twenty-20 matches are too few to come to any conclusion and are not included in the current study.

*(b) Left-handers in Cricket history*

We analyzed the handedness distribution for batsmen and bowlers in Test cricket and ODI cricket. For Batsmen the data for year-end-rankings are collected from 1877 through 2011 for Test cricket, and between 1971 and 2011 in ODI cricket. Data for rankings is accessed from ICC's open source online database. In Test cricket we collect the information of *1553* bowlers, *131* bowlers bowled left-handed. In ODI cricket, *85* out of *1234* bowlers are left-handed. For batsmen, in Test cricket *482* players out of *2611* bat left-handed, while in ODI cricket *384* players of a total of *1880* bat left-handed.

In Figure 1 we temporally track the handedness distribution in order to understand the impact of handedness on expert performance. We track the ICC points of every player over time and then proceed to study the effect of left-handedness and leadership in Cricket. We observe that in Test matches (1877-2011) and in ODI matches (1971-2011) among the top 100 batsmen (as measured by the ICC points), number of left-handers increases according to a second-order polynomial (Figure 1A). In both forms of cricket, we observe a slow increase in number of left-handed bowlers who are among the top 100 in year-end ICC rankings (Figure 1B).

Next we focus our attention on the fraction of left-handers among the top 100 performers in batting and bowling in all forms of Cricket (Figure 1C-1F). We evaluated the fraction of left-handers present in every ranking interval for top 100 performers. If playing left-handed provides an advantage in performance, we expect fraction of left-handers to occupy higher rankings and that the fraction of left-handers would decrease with ranking intervals. There exists a higher fraction of left-handers occupying higher ranks representing top batting performance in Test cricket (Figure 1C) and ODI cricket (Figure 1D). However, we don't observe any relation between fraction of left-handed bowlers and ranking intervals. This indicates that for bowlers, bowling left-handed is not an advantage in any forms of Cricket (Figure 1E – 1F).

## 3. RESULTS AND DISCUSSION

We proceed further to investigate whether captain's left-handedness prove beneficial in batting and bowling performances of players. We analyze the association between batting-handedness of the captain on the number of wickets claimed by bowler and number of runs scored by batsmen in ODI and Test cricket. In Figure 2 we graphically show the results found in our analysis - in both forms of Cricket, players perform significantly better under left-handed captains as compared to performance under right-handed captains. In ODI cricket, on an average bowlers claim *18* wickets while playing under left-handed captains, almost *1.4* times than what claimed while playing under right-handed captains. Examining the batting performance of batsmen, we observe that in ODI cricket batsmen scored *343* runs under left-handed captaincy, almost *1.4* times than what scored under right-handed captaincy. It is to be noted that in cricket bowlers also get to score runs and batsmen are used as part-time bowlers to dismiss opponent batsmen. Again, a bowler is judged by his bowling performance (i.e. number of wickets claimed) and batsman is judged by his batting performance (i.e. runs scored). This results in the data getting skewed. We therefore separate the batting performance of batsmen from batting performance of bowlers by considering the elite performances - being in the top *5%* (or top *25%*) of batting and bowling performance. Figure 3 indicates that elite performance (top 5% of runs scored or wickets claimed) under left-handed captains is significantly raised net of team fixed effects.

Regression analyses incorporating team fixed effects confirm the predictive power of laterality of captains on individual performance of players. To test the sensitivity of the above observations and analyze the association of laterality of leaders with player's performance, we conduct logistic regressions of the form

$$\Pr(h_i) = f\left(\alpha D^l_{\ i} + \sum_y \beta_y Team_{yi}\right) \qquad (1)$$

Where $D^l_{\ i}$ is a dummy variable, which takes value 1 if a captain *i* bats left-handed and is 0 otherwise, $h_i$ is an indicator variable of performance of a player (under leadership of captain *i*) being in the elite performers. In order to eliminate bias due to omitted variables, the logistic regression includes a full set of fixed effects for each of teams, where the indicator variables [(Team)]$_{yi}$ ∈ {0, 1} are equal to *1* if the captain *i* leads team *y*. We define elite performers as those who belong to the top *5%* of runs scored (Table 2) or top *25%* of wickets claimed (Table 3). As confirmatory robustness check we perform our analyses with different definitions of elite performance - being in the top *1%* or top *10%* or top *25%* of batting and bowling performance respectively. Our results are also consistent with a linear model, which takes the form

$$P_i = f(\alpha D^l_i + \sum_y \beta_y Team_{yi}) \qquad (2)$$

Where $P_i$ is the performance of a player (batsman or bowler) under the leadership of captain *i*. The number of runs scored judges the batsmen, while the number of wickets claimed judges the bowlers (See Table 4-5). These results suggest that left-handed captains are more successful than their right-handed counterparts.

Next, we empirically investigate whether left-handedness is positively associated with the outcome of a contest. We hypothesize that being left-handed provides an advantage to the success of the captain and has a stronger effect on the outcome of the match as compared to the effect of strength of the team. The strength of a team is measured by the median batting average and median bowling average of all players in the team. A team with high median batting average is considered stronger compared to team with low median batting average. Similarly a team with low median bowling average is stronger than a team with high median batting average. To assess the association between left-handedness of the captain and the team performance, we perform a linear regression of the form

$$RR_i = f(\alpha_1 D^l_i + \alpha_2 B^1_{Avg,i} + \alpha_3 B^2_{Avg,i} + \sum_y \beta_y Team_{yi}) \qquad (3)$$

Where $D^l_i$ is a dummy variable which takes value *1* if captain *i* bats left-handed and is *0* otherwise, $RR_i$ is the *run-rate* of a team. We choose *run-rate* as an indicator of match outcome, since in Test cricket, there are situations in which the outcome of contest remains undecided (known as "match drawn"). In a contest, a team with higher *run-rate* is the better performer than the team with lower *run-rate*. The control variable $B^1_{Avg,i}$ is the median batting average and $B^2_{Avg,i}$ is the median bowling average of a team, led by captain *i*. The regression includes a full set of fixed effects for each of captains, where the indicator variables [(Team)]$_{yi}$ ∈ {0, 1} are equal to *1* if the captain *i* leads team *y*. We summarize our results in Table 6.

Examining the interface of team performance, and left-handedness of the captain shows an intricate relationship. In ODI and Test cricket, left-handedness ($D^l$) of the captain is significantly (***$p < 0.00001$*) and positively associated with the team performance ($RR_i$), as indicated by the positive coefficient (Table 6). We use standard BIC statistics to interpret the improvement in fit of the regression model when $D^l$ is added to the control variable model. A drop in the BIC statistic of *2-6* provides "strong" evidence that the variables added to the model improve the fit of the overall model (Raftery 1995). The above pattern suggests that, left-handed captains are more successful than their right-handed counterparts.

As supported by the second-order polynomial fit, in the past couple of decades the number of top left-handed batsmen has swelled beyond comparison with that of previous generations (Figure 1). We observed a steady rise in number of elite left-handed batsmen in both forms of Cricket. It remains to be seen whether the number rises in the future and subsequently reach a steady value. With the increase in number of elite left-handed batsmen, we have witnessed the successful leadership skills of left-handed Cricket captains like Clive Lloyd, Graeme Smith, Sourav Ganguly and Stephen Fleming.

Earlier models on laterality suggest that left-handers enjoy an advantage because they are a minority – less than *10%* of the total population being left-handed (Abrams & Panaggio 2012; Ghirlanda, Frasnelli & Vallortigara 2009). In other words, the advantage arises from the minority and not from being left-handed. In this work, we suggest that the left-handers enjoy a leadership advantage, presumably distinct from the minority advantage. We focused on the most important aspect of the game - leadership. In Cricket an outcome of a game depends a lot on the leadership skills of the captains. As mentioned earlier the captain takes on-field decisions like bowling changes, field placements that affect the outcome of a game. We explored the idea whether left-handed captains have a tactical or strategic advantage over their right-handed rivals in Cricket. We found evidence to suggest that left-handed captains are more successful in extracting better performance from the batsmen and bowlers than the right-handed captains.

Future work remains to examine mechanisms that underlie the superior performance of left-handed captains in cricket. One possible explanation for the success of left-handed captains in ODI cricket could be imbedded in inherent ability of left-handers to think outside the box in result oriented competitions. This is backed by a recent study that suggest; left-handers are quicker in responding to specific tests than right-handers (Cherbuin & Brinkman 2006). Another conjecture could be the innate ability of left-handed individuals to tackle the pressure of leading a side, which might have more to do with neurological advantage (Knecht et al., 2000; Denny & O'Sullivan, 2007; Christopher, Roberts, Tim, Stuart, & M, 2009). One could argue that the strategic advantages enjoyed by left-handed captains arise solely by the fact that other captains are not accustomed to the strategies. It is to be noted that strategies designed by captains mostly include team-selection, selecting batting order, bowling changes and field placements. In terms of field placements of *11* players, specific fielding positions are assigned to the players. A captain could at most, change the fielder from one position to another but not innovate a completely new fielding position (See *http://static.espncricinfo.com/db/ABOUT_CRICKET/fielding-positions.pdf* for more details). However, further studies are required to understand how a captain reacts to an adverse situation. For example, if batsmen of opponent team are scoring at a high rate or bowlers of opponent team are performing better, how fast does a left-handed captain adapt to the situation compared to right-handed captains. Such an analysis is reserved for future studies, which would include qualitative aspects of watching every proceedings of the game. Further studies also include how strength of a team could influence the performance of a player in the team. To the best of our knowledge cricket is essentially independent of other team member's performance and there is no study, which has demonstrated the "interdependence" of players' performance in Cricket.

There are few more aspects, which deserve closer scrutiny. First, it has been seen earlier that left-handedness is more pronounced in men than women in many professional sports like tennis (Loffling, Hagemann, & Strauss 2012). This is again attributed to the common gender differences observed in general population. These findings are however restricted to activities not involving any team involvement (Loffling, Hagemann, & Strauss 2012). Sport-specific performance and leadership abilities depending on gender thus remains an open idea for further research. Secondly, in most cricketing nations there are many domestic competitions that serve as a road to international exposure. Most of the domestic players fail to make any international appearance due to lack of skills needed at the elite level. Due to non-availability of consistent information on domestic games we are unable to study the number of left-handers with high performance level and leadership skills in cohort of domestic players. We think that even in domestic level, playing left-handed will provide an advantage of being selected at international level.

There are certain limitations in our study that needs to be addressed. In this paper we have addressed those captains who bat left-handed. However, recent works indicate that preference for left-handedness in sports related tasks doesn't necessarily imply natural left-handedness (Loffling, Solter & Hagemann 2014). In the history of cricket, there are instances of players who played left-handed but in normal life they were right handed. Examples include West Indies' Brian Lara, India's Sourav Ganguly and Australia's Mark Taylor who batted left-handed but used their right hand in routine life (Meher 2012). However such situations – right-handed in normal life and left-handed in sports and vice-versa exist in many other sports as well (Loffling, Solter & Hagemann 2014). Examples include Ben Hogan, Arnold Palmer and Nick Price in golf - naturally left-handed but played right-handed. Again, Phil Mickelson is right-handed naturally but plays left-handed. The mechanism that helps left-handed leaders to be successful is subject to further analysis and is beyond the current models of handedness in competitive sports. Another potential limitation is regarding the work environment under left-handed captains. For example, do left-handed captains provide a better environment for performance and there is lesser conflict among the team members? Answering these questions need further studies involving qualitative techniques.

To summarize, we have presented to the best of our knowledge, the first evidence of left-handedness and leadership advantage in the context of interactive contests. Our analysis shows that left-handedness has a positive effect on leadership and left-handed captains are more successful in educing better performance from the players as compared to right-handed captains, indicating that they are able to handle the pressure of leadership better. While our findings cannot establish a causal relationship, it nonetheless strongly suggests the positive relation between captain's laterality and individual performance, and provides insight into the mechanisms by which leader's left-handedness may influence performance.

However there is also a need to extend our study to other fields involving strong leadership - the role of political leadership in team performance and whether left-handed leaders are able to bring significant economic growth in a nation compared to right-handed counterparts. Leadership is critical for successful management of teams. Presence of an individual with exceptional entrepreneurial and motivational skills in any teamwork is essential. Community leaders who are guided by collective interests provide resilience to change in governance (Guiterez, Hilborn, & Defeo 2011). While our data is not a representative sample of world population of left-handers, our research indicates that perhaps left-handed leaders could be more influential in providing that coveted resilience and motivation to a team.

## REFERENCES

Abrams, D. M., & Panaggio, M. J.  2012 A model balancing cooperation and competition can explain our right-handed world and the dominance of left-handed athletes. *The Journal of the Royal Society Interface*, 9, 2718-2722.

Aggleton, J. P., Bland, J. M., & Kentridge, N. J., R. W.and Neave. 1994 Handedness and longevity : archival study of cricketers. *British Medical Journal*, 309, 1681–1684.

Baker, J., & Schorer, J. 2013 The southpaw advantage? – Lateral preference in mixed martial arts. *PLoS ONE*, 8(11): e79793

Beratis, I. N., Rabavilas, A. D., Kyprianou, M., Papadimitriou G. N. & Papageorgiou C. 2013 Investigation of the link between higher order cognitive functions and handedness. *J Clin Exp Neuropsychol.,* 35(4), 393-403

Brooks, R., F., B. L., Jennions, M. D., & Hunt, J. 2004 Sinister strategies succeed at the cricket world cup. *Proceedings of Royal Society B* (Suppl. 3), 271, S64–S66.

Cherbuin, N., & Brinkman, C. 2006 Hemispheric interactions are different in left-handed individuals. *Neuropsychology*, 20, 700–707.

Christopher, P. C., Roberts, N., Tim, J. C., Stuart, J. L., & M, G.-F. 2009 The effect of handedness on academic ability: A multivariate linear mixed model approach. *Laterality: Asymmetries of Body, Brain and Cognition*, 15, 451–464.

Coren, S. 1995 Differences in divergent thinking as a function of handedness and sex. *Am J Psychol,* 108 (3), 311–325.

Cricinfo. 2014 Retrieved 02-03-2014, from: www.espncricinfo.com

Denny, K., & O'Sullivan, V. 2007 The economic consequences of being left-handed: some sinister results. *Journal of Human Resources*, 42, 353–374.

Duch, J., Waitzman, J. S., & Amaral, L. A. N.  2010 Quantifying the performance of individual players in a team activity. *PLoS ONE*, 5, e10937.

Faurie, C., & Raymond, M. 2004 Handedness frequency over more than ten thousand years. *Proceedings of Royal Society B* (Suppl. 3), 271, S43–S45.

Goldstein, S. R., Young, C. A. 1996 Evolutionary stable strategy of handedness in major league baseball. *J Comp Psych,* 110, 164 - 169.

Ghirlanda, S., Frasnelli, E., & Vallortigara, G. 2009 Intraspecific competition and coordination in the evolution of lateralization. *Phil. Trans. R. Soc. B*, 364, 861 – 866.

Ghirlanda, S., & Vallortigara, G. 2004 The evolution of brain lateralization: a game-theoretical analysis of population structure, *Proc Biol Sci*, 271(1541), 853 - 857

Gursoy, R. 2008 Effects of left- or right-handed preference on the success of boxers in Turkey. *Br J sports Med*, 43, 142-144

Gutierez, N. L., Hilborn, R., & Defeo, O. 2011 Leadership, social capital and incentives promote successful fisheries. *Nature*, 470, 386–389.

Harris, L. J. 2010 In fencing, what gives left-handers the edge? Views from the present and the distant past. *Laterality,* 15, 15-55.

Holder, M. K. 2001 Why are more people right-handed? Retrieved on 11-03-2013, http://www.scientificamerican.com/article/why-are-more-people-right/

Holtzen, D. W. 2000 Handedness and professional tennis. *Int J Neurosci,* 105, 101 - 119

Knecht, S., Drager, B., Deppe, M., Bobe, L., Lohmann, H., Floel, A., et al. 2000 Handedness and hemispheric language dominance in healthy humans. *Brain*, 123, 2512–2518.

Loffling, F., Hagemann, N., & Strauss, B. 2012 Left-handedness in professional and amateur tennis. *PLoS ONE*, 7, e49325.

Loffling, F., Solter, F., & Hagemann, N. 2014 Left preference for sport task does not necessarily indicate left-handedness: Sport-specific lateral preferences, relationship with handedness and implications for laterality research in behavioral sciences. *PLoS ONE*, 9, e105800

Meher, H. K. 2012 The weird world of ambidextrous test cricketers. Retrieved on 11-03-2013, http://www.theroar.com.au/2012/03/29/the-weird-world-of-ambidextrous-test-cricketers/.

Mukherjee, S. 2012 Identifying the greatest team and captain - a complex network approach to cricket matches. *Physica A*, 391, 6066–6076.

Radicchi, F. 2005 Universality, limits and predictability of gold medal performances at the olympic games. *PLoS ONE*, 7, e40335.

Raftery A. E. 1995 Bayesian model selection in social research. *Sociological Methodology* 25, 111-163

Raymond, M., Pointer, D., Dufour, A. B., & Moller, A. P. 1996 Frequency-dependent maintenance of left-handedness in humans. *Proceedings of Royal Society B*, 263, 1627–1633.

Saavedra, S., Mukherjee, S., & Bagrow, J. P. 2012 Is coaching experience associated with effective use of timeouts in basketball ? *Scientific Reports*, 2, 676-679.

Saavedra, S., Powers, S., McCotter, T., Porter, M. A., & Mucha, P. J. 2009 Mutually antagonistic interactions in baseball networks. *Physica A*, 389, 1131–1141.

Skinner, B. 2010 The price of anarchy in basketball. *Journal of Quantitative Analysis in Sports*, 6,3

Tommasi, L. 2009 Mechanisms and functions of brain and behaviorial asymmetries. *Phil Trans R Soc B,* 364, 855-589

Vallortigara, G. 2006 The evolutionary psychology of left and right: costs and benefits of lateralization, *Dev Psychobiol*, 48(6), 418 - 427

Wood, C. J., & Aggleton, J. P. 1989 Handedness in fast ball sports: do left-handers have an innate advantage ? *British Journal of Psychology*, 80, 227–240

Ziyagil, M. A., & Gursoy, R., Dane, S., & Yuksel, R. 2010 Left-handed wrestlers and more successful. *Perceptual and Motor Skills*, 111, 65-70

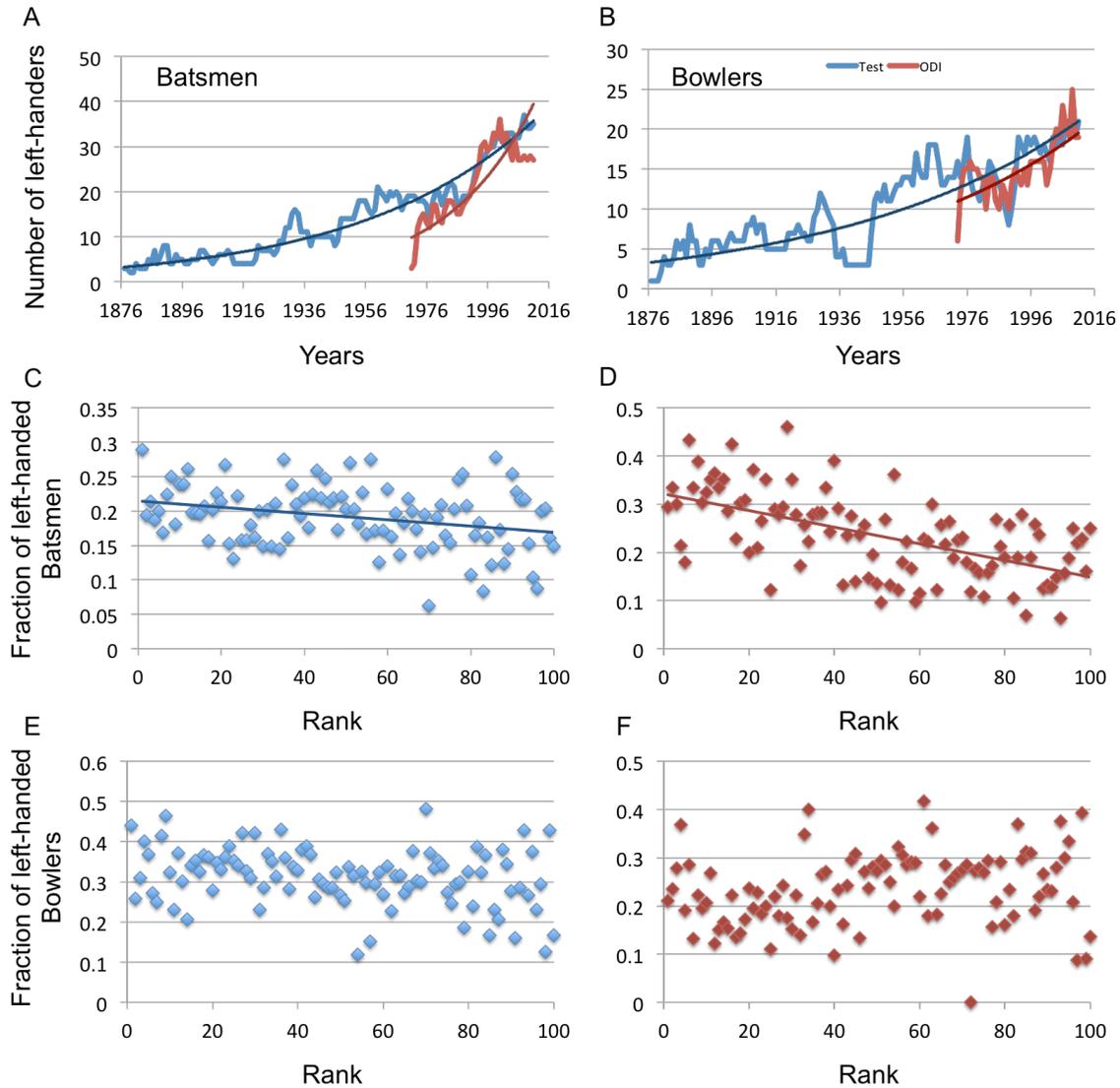

Figure 1. In (A) we plot the number of left-handed batsmen in the top 100 ICC rankings over time for Test cricket (1877-2011) and ODI cricket (1971-2011). The data is fitted to a second-order polynomial ($R^2 = 0.9225$, $p < 10^{-07}$ for Test cricket and $R^2 = 0.8158$, $p < 10^{-07}$ for ODI cricket) indicating an increasing trend in number of left-handed batsmen in both forms of Cricket. (B) We observe a similar increasing trend in number of left-handed bowlers in top 100 ICC rankings over time in Test cricket ($R^2 = 0.8217$, $p < 10^{-07}$) and ODI cricket ($R^2 = 0.8543$, $p < 10^{-07}$). We plot the fraction of left-handers at various ranking intervals of top 100 batsmen in (C) Test cricket (1877-2011) and (D) ODI cricket (1971-2011). The fraction of left-handers decrease with ranking intervals as shown by the linear fit in C ($R^2 = 0.0864$, $p = 0.003$) and D ($R^2 = 0.3363$, $p < 10^{-07}$). There is no clear pattern of relationship between fraction of left-handed bowlers and ranking intervals (E and F).

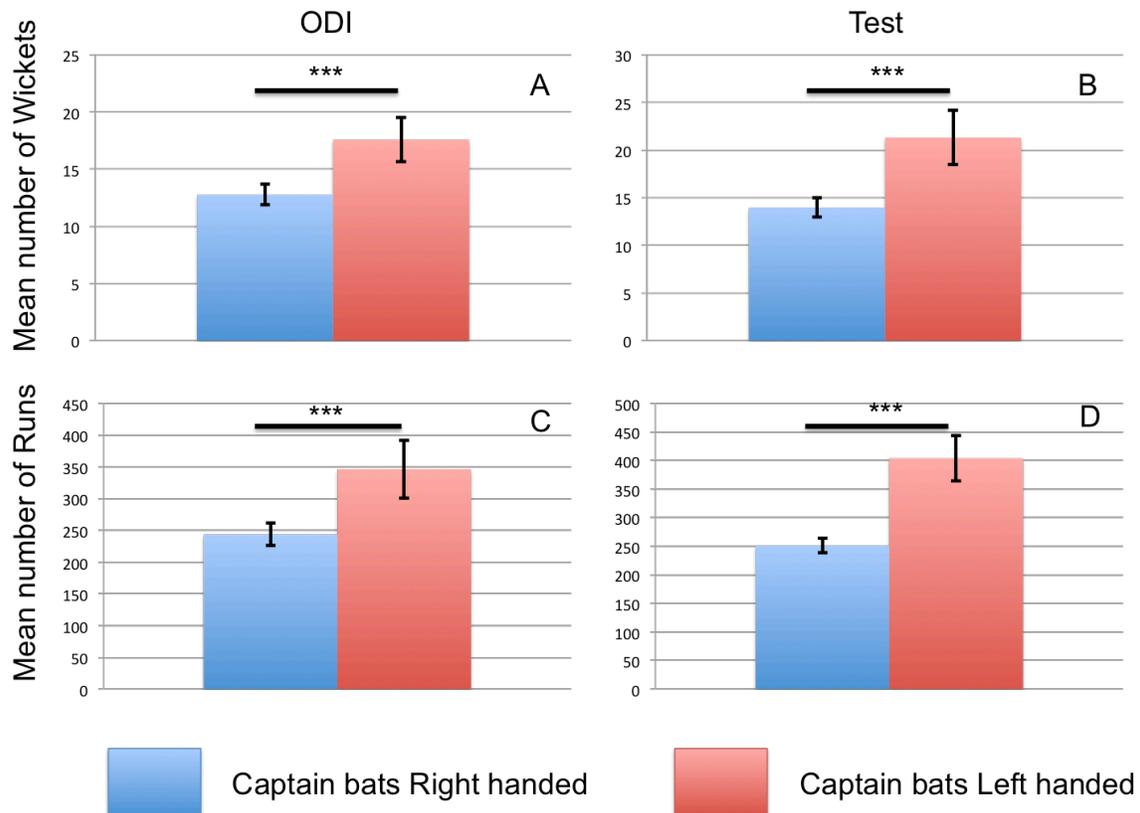

Figure 2. Figure demonstrates the batting and bowling performance of players under right-handed and left-handed captains. (A) In ODIs, between 1971 and 2013 on an average bowler took *18* wickets under left-handed captains, significantly higher than that under right-handed leaders. (B) In Test cricket (1877-2013) as well, left-handed captains show significantly better success rate in terms of bowling performance of bowlers who played under their leadership. Similarly, in (C) ODIs and (D) Test cricket, batsmen scored significantly more runs on an average under left-handed captains than under right-handed captains. Here, bars represent *95%* confidence intervals obtained by resampling, *** denotes *p < 0.001*.

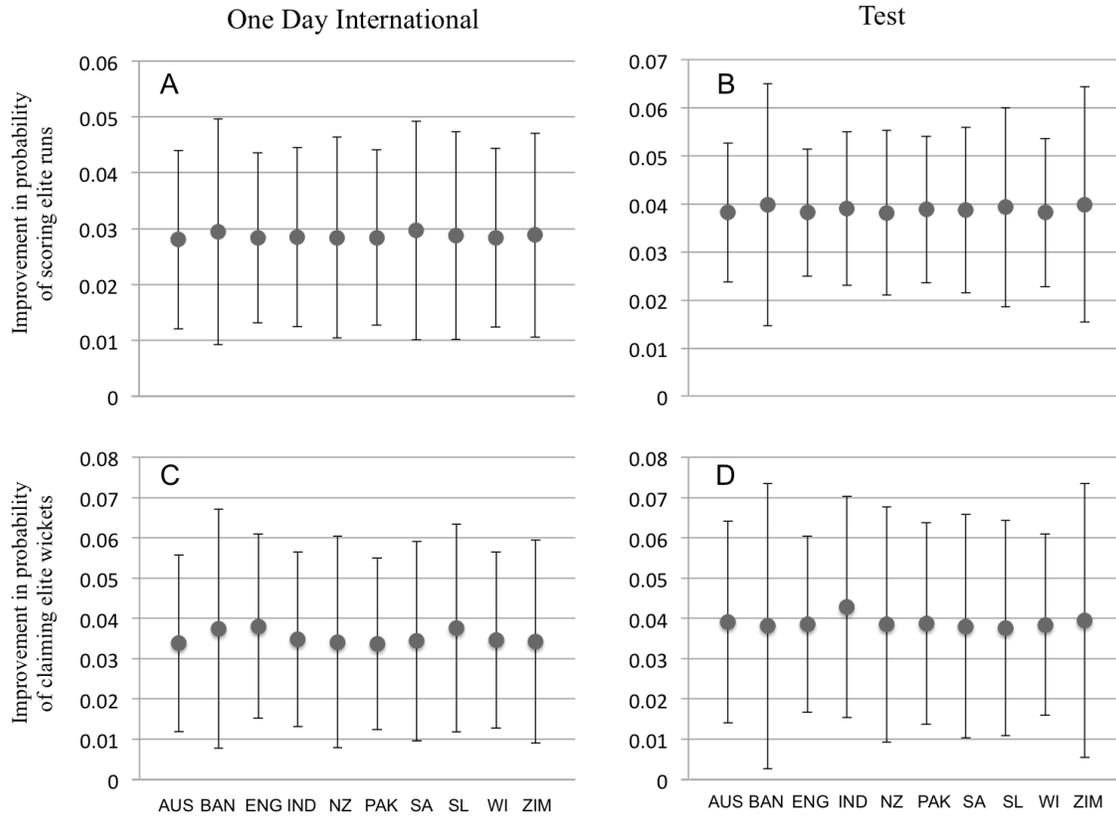

Figure 3. Individual performance of batsmen and bowlers improves under left-handed captains. All players belong to different teams show an approximately *4%* on average statistically significant improvement in their chances of elite performance when playing under leadership of left-handed captains (*95%* confidence intervals shown). We observe no significant dominance of particular team, indicating that there is no team effect in left-handedness and leadership.

| Name | Team | Career Span | Career Batting Average | Significant contribution (Source: Cricinfo) |
|---|---|---|---|---|
| Clem Hill | Australia | *1896 - 1912* | *39.21* | Ranked best batsman according to ICC points in *1902*; First player to reach *800* ICC points |
| Maurice Leyland | England | *1928 - 1938* | *46.06* | Ranked 4th best batsman according to ICC points in *1936* |
| Eddie Paynter | England | *1931 - 1939* | *59.23* | Ranked 5th best batsman according to ICC points in *1939*; Averaged *84.42* for his seven Test matches against Australia, a figure no other English players achieved |
| Arthur Morris | Australia | *1946 - 1955* | *46.48* | Ranked 2nd best batsman according to ICC points in *1950* |
| Neil Harvey | Australia | *1948 - 1963* | *48.41* | Ranked best batsman according to ICC points in *1953* |
| Bert Sutcliffe | New Zealand | *1947 - 1965* | *40.10* | Ranked 4th best batsman according to ICC points in *1955* |
| Sir Gary Sobers | West Indies | *1954 - 1974* | *57.78* | Excellent batsman, bowler and enterprising captain; Knighted in 1975 for his contributions to cricket |

Table 1. Information and cricketing achievements of left-handed batsmen during 1900s through 1960s

| | One Day International (1971 – 2013) β (SE) | | Test (1877 -2013) β (SE) | |
|---|---|---|---|---|
| $D^l$ | 0.638***(0.137) | 0.620***(0.142) | 0.783***(0.116) | 0.732***(0.122) |
| Fixed Effects | | | | |
| *Team* | | Y | | Y |
| Observations | 4717 | 4717 | 6965 | 6774 |

Table 2. Results for the logistic regression used for predicting the effect of left-handedness of captains on individual performance of batsmen (*h*) including teams fixed effects. Elite performers are those who belong to the top *5%* of runs scored. The results are consistent with top *1%* or top *10%* or top *25%* of runs scored; *** $p < 0.001$

|  | One Day International (1971 – 2013) β (SE) | | Test (1877 -2013) β (SE) | |
| --- | --- | --- | --- | --- |
| $D^l$ | 0.735***(0.173) | 0.806***(0.176) | 0.721***(0.179) | 0.784***(0.191) |
| Fixed Effects | | | | |
| *Team* | | Y | | Y |
| Observations | 2764 | 2764 | 2625 | 2526 |

Table 3. Results for the logistic regression used for predicting the effect of left-handedness of captains on individual performance of bowlers (*h*) including teams fixed effects. Elite performers are those who belong to the top *5%* of wickets claimed. The results are consistent with top *1%* or top *10%* or top *25%* of wickets claimed; *** $p < 0.001$

|  | One Day International (1971 – 2013) β (SE) | | Test (1877 -2013) β (SE) | |
| --- | --- | --- | --- | --- |
| $D^l$ | 102.866***(23.857) | 99.659***(24.089) | 152.650***(22.583) | 146.712***(23.414) |
| Fixed Effects | | | | |
| *Team* | | Y | | Y |
| Observations | 4717 | 4717 | 6965 | 6965 |
| $R^2$ | 0.005 | 0.018 | 0.0121 | 0.0198 |

Table 4. Results for the linear model used for predicting the effect of left-handedness of captains on number of runs scored by players including teams fixed effects. The results are consistent with elite performance of runs scored; *** $p < 0.001$

|  | One Day International (1971 – 2013) β (SE) |  | Test (1877 -2013) β (SE) |  |
|---|---|---|---|---|
| $D^l$ | 4.792***(1.076) | 4.861***(1.088) | 7.348***(1.552) | 7.403***(1.621) |
| Fixed Effects |  |  |  |  |
| Team |  | Y |  | Y |
| Observations | 2764 | 2764 | 2625 | 2625 |
| $R^2$ | 0.0089 | 0.025 | 0.0128 | 0.03 |

Table 5. Results for the linear model used for predicting the effect of left-handedness of captains on number of wickets claimed by players including teams fixed effects. The results are consistent with elite performance of wickets claimed; *** $p < 0.001$

|  | One Day International (1971 – 2013) β (SE) |  | Test (1877 -2013) β (SE) |  |
|---|---|---|---|---|
| $B^1_{AVG}$ | 0.057***(0.002) | 0.057***(0.002) | 0.028***(0.001) | 0.027***(0.001) |
| $B^2_{AVG}$ | 0.006***(0.001) | 0.006***(0.001) | -0.001(0.001) | -0.001(0.001) |
| $D^l$ |  | 0.098***(0.027) |  | 0.096***(0.024) |
| Fixed Effects |  |  |  |  |
| Team | Y | Y | Y | Y |
| Observations | 5363 | 5363 | 3970 | 3970 |
| BIC | 14205.2 | 14200.3 | 8268.678 | 8262.186 |

Table 6. Results for the linear regression used for predicting the effect of left-handedness ($D^l$) of captains on outcome of a match ($RR$) controlling for median batting average of team ($B^1_{AVG}$), median bowling average of team ($B^2_{AVG}$), and team fixed effects. We observe the stability of the coefficient estimates of $D^l$ when the regression is run on $D^l$ only vs. the full model and that the BIC statistics indicate "strong" support that a model specified with $D^l$ significantly increases the explanatory power of the model over the control variable model. ; *** $p < 0.001$

# APPENDIX

**Glossary of Cricket Terms**

Runs scored – Essentially it refers to the scoring by a batsman. A single run by a batsman adds up to the total score of a team. Thus "a team's score is 100" is referred to as "team has scored 100 runs".

Run-rate – It is defined as the number of runs scored by a team divided by the number of overs played. Overs refer to the number of balls bowled. 1 Over = 6 balls bowled. Thus if a team scores 300 runs in 50 overs, the run rate is 300/50 = 6.00; while another team scoring 250 runs in 45 overs will have a run-rate of 5.56

Wickets claimed – This refers to the dismissal of a *batsman* by a *bowler* in the Cricket ground. A bowler claimed *3* wickets means he dismissed *3* batsmen.

Batting Average – Number of runs scored by a batsman divided by the number of times he has been dismissed.

Bowling Average – Number of runs conceded by a bowler divided by the number of wickets claimed by the bowler.

# ACKNOWLEDGEMENT

The author thanks Cricinfo for public availability of Cricket data. The author also thanks the two unknown referees for their valuable insights and comments, which has improved the paper substantially.